# Pressure-Induced Superconductivity and Structural Phase Transitions in Magnetic Topological Insulator Candidate MnSb$_4$Te$_7$

Cuiying Pei[1#], Ming Xi[2#], Qi Wang[1,3], Wujun Shi[4,5], Lingling Gao[1], Yi Zhao[1], Shangjie Tian[2], Weizheng Cao[1], Changhua Li[1], Mingxin Zhang[1], Shihao Zhu[1], Yulin Chen[1,3,6], Hechang Lei[2*], and Yanpeng Qi[1,3,7*]

1. School of Physical Science and Technology, ShanghaiTech University, Shanghai 201210, China
2. Department of Physics and Beijing Key Laboratory of Opto-electronic Functional Materials & Micro-nano Devices, Renmin University of China, Beijing 100872, China
3. ShanghaiTech Laboratory for Topological Physics, ShanghaiTech University, Shanghai 201210, China
4. Center for Transformative Science, ShanghaiTech University, Shanghai 201210, China
5. Shanghai High Repetition Rate XFEL and Extreme Light Facility (SHINE), ShanghaiTech University, Shanghai 201210, China
6. Department of Physics, Clarendon Laboratory, University of Oxford, Parks Road, Oxford OX1 3PU, UK
7. Shanghai Key Laboratory of High-resolution Electron Microscopy, ShanghaiTech University, Shanghai 201210, China

# These authors contributed to this work equally.
* Correspondence should be addressed to Y.P.Q. (qiyp@shanghaitech.edu.cn) or H.C.L. (hlei@ruc.edu.cn)

**ABSTRACT:**

**The magnetic van der Waals crystals (MnX$_2$Te$_4$)$_m$(X$_2$Te$_3$)$_n$ (X = Sb, Bi) have drawn significant attention due to their rich topological properties and the tenability by external magnetic field. In this work, we report on the discovery of superconductivity in magnetic topological insulator candidate MnSb$_4$Te$_7$ ($m$ = 1, $n$ = 1) via the application of high pressure. The antiferromagnetic ordering is robust to pressure until 8 GPa and then fully suppressed. The carrier type converts from hole- to electron-type accompanied with structural phase transition at around 15 GPa. Superconductivity emerges near the critical pressure 30 GPa where MnSb$_4$Te$_7$ converted into a simple cubic phase. Interestingly, MnSb$_4$Te$_7$ shows a dome-like phase diagram with a maximum $T_c$ of 2.2 K at 50.7 GPa. The results**

**demonstrate that MnSb$_4$Te$_7$ with nontrivial topology of electronic states display new ground states upon compression.**

INTRODUCTION

Superconductivity and topological quantum state are two important fields of frontier research in condensed matter physics. Since the discovery of topological insulators (TIs), extensive investigations have been conducted for observing topological superconductivity (TSC) and Majorana fermions, which have potential applications to fault-tolerant topological quantum computation.[1-4] One feasible route to realize topological superconductor is to search for intrinsic superconductivity in topological materials by doping/intercalation or applying high pressure.[5-18] Indeed, superconductivity achieved in this manner has been observed in typical topological insulators. For example, topological insulator Bi$_2$Se$_3$ can be induced to become a bulk superconductor, by copper intercalation in the van der Waals gaps between the Bi$_2$Se$_3$ layers or application of pressure above 11 GPa.[5, 19]

When magnetic ordering is introduced to topological insulators, such intrinsic magnetic topological insulators (MTIs) can display a lot of unusual physical properties, such as Chern insulator state with chiral edge state hosting quantum anomalous Hall effect and axion insulator state with axion electrodynamics[20-23]. Recently, intrinsic MTIs of (MnBi$_2$Te$_4$)$_m$(Bi$_2$Te$_3$)$_n$ has been theoretically predicted and experimentally synthesized to have tunable magnetic properties and topologically nontrivial surface states[22-36]. (MnBi$_2$Te$_4$)$_m$(Bi$_2$Te$_3$)$_n$ crystallizes in a typical van der Waals layered structure, sharing a similar crystal structure with Bi$_2$Te$_3$, a typical TI under ambient conditions. Although the topological properties of these materials have been intensively studied in the past few years, no superconductivity has been observed either by chemical doping or apply pressure.

Compared with (MnBi$_2$Te$_4$)$_m$(Bi$_2$Te$_3$)$_n$, the (MnSb$_2$Te$_4$)$_m$(Sb$_2$Te$_3$)$_n$ family remains much less explored. In particular, MnSb$_4$Te$_7$ ($m$ = 1, $n$ = 1), one member of the (MnSb$_2$Te$_4$)$_m$(Sb$_2$Te$_3$)$_n$ family, is a magnetic topological system with versatile topological phases that can be manipulated by both carrier doping and magnetic field[37].

In this work, we report on the discovery of superconductivity on intrinsic MTI candidate MnSb$_4$Te$_7$ via the application of high pressure. Application of pressure effectively tunes physical properties and crystal structure of MnSb$_4$Te$_7$. Pressure-induced two high-pressure phase transitions are revealed in MnSb$_4$Te$_7$, and superconductivity with a dome shape behavior is observed after MnSb$_4$Te$_7$ converts into a simple cubic phase.

**EXPERIMENTAL SECTION**

**Sample preparation.** Sample synthesis, structural and composition characterizations. Single crystals of MnSb$_4$Te$_7$ were grown using the self-flux method. The starting materials Mn (piece, 99.99%), Sb (grain, 99.9999%), and Te (lump, 99.9999%) were mixed in an Ar-filled glove box at a molar radio of Mn : Sb : Te = 1 : 11.5 : 15. The mixture was placed in an alumina crucible, which was then sealed in an evacuated quartz tube. The tube was heated to 1023 K for 24 h and kept at that temperature for 20 h. Then, the tube was slowly cooled down to 881 K at a rate of 0.5 K/h followed by separating the crystals from the flux by centrifuging. Finally, the ampoule was taken out from the furnace and decanted with a centrifuge to separate MnSb$_4$Te$_7$ single crystals from the flux. MnSb$_4$Te$_7$ single crystals are stable in the air. The phase and quality examinations of MnSb$_4$Te$_7$ single crystal were performed on the Bruker AXS D8 Advance powder crystal x-ray diffractometer with Cu $K_{\alpha1}$ ($\lambda$ = 1.54178 Å) at room temperature. Magnetotransport measurements were performed on Physical Property Measurement System (PPMS). The magnetization measurement was carried on a Magnetic Property Measurement System (MPMS).

**High pressure measurements.** High pressures were generated with diamond anvil cell (DAC) as described elsewhere[38-41]. *In situ* high-pressure x-ray diffraction (XRD) measurements were performed at beamline BL15U of Shanghai Synchrotron Radiation Facility (x-ray wavelength $\lambda$ = 0.6199 Å). A symmetric DAC with 200 μm culet was used with rhenium gasket. Silicon oil was used as the pressure transmitting medium (PTM) and pressure was determined by the ruby luminescence method[42]. The two-dimensional diffraction images were analyzed using the FIT2D software[43]. Rietveld refinements on crystal structures under high pressure were performed using the General

Structure Analysis System (GSAS) and the graphical user interface EXPGUI[44-45]. An *in situ* high-pressure Raman spectroscopy investigation on MnSb$_4$Te$_7$ was performed by a Raman spectrometer (Renishaw in-Via, UK) with a laser excitation wavelength of 532 nm and low-wavenumber filter. A symmetry DAC with 200 μm culet was used, with silicon oil as the PTM. *In situ* high-pressure resistivity and Hall effect measurements of MnSb$_4$Te$_7$ were conducted on a nonmagnetic DAC. A piece of nonmagnetic BeCu was used as the gasket. Cubic BN/epoxy mixture layer was inserted between BeCu gasket and electrical leads as insulator layer. Four Pt foils were arranged according to the van der Pauw method. To monitor the evolution of $T_N$, NaCl was used as PTM to get a quasi-hydrostatic pressure to the sample.

**RESULTS AND DISCUSSION**

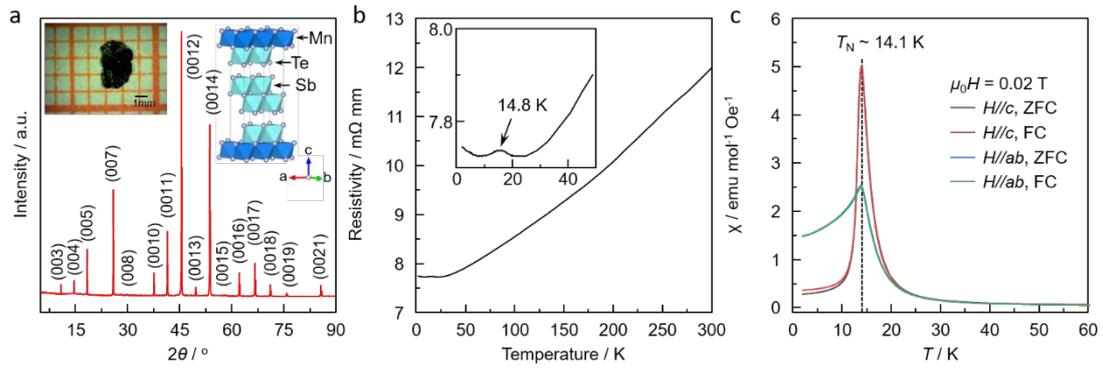

Figure 1. (a) The room temperature x-ray diffraction peaks from the *ab* plane of MnSb$_4$Te$_7$ crystal. Inset: Image of a typical MnSb$_4$Te$_7$ single crystal synthesized in this work and the schematic crystal structure of MnSb$_4$Te$_7$; (b) Temperature dependent of *ab*-plane resistivity $\rho(T)$ at zero magnetic field. Insets show the enlarged view of the $\rho(T)$ curve from 0 to 50 K; (c) The temperature dependent magnetic susceptibility χ under $\mu_0H$ = 0.02 T for *H//c* (black and red lines) and *H//ab* plane (blue and green lines).

At ambient pressure, MnSb$_4$Te$_7$ adopts a rhombohedral $P\bar{3}m1$ structure with alternate stacking of one MnSb$_2$Te$_4$ septuple layers (SLs) and one Sb$_2$Te$_3$ quintuple layer (QLs), as shown in the right inset of Figure 1a. All of peaks in the XRD pattern of a crystal can be well indexed by the (00*l*) reflections of MnSb$_4$Te$_7$ (Figure 1a), indicating that the crystal surface is parallel to the *ab* plane and perpendicular to the *c* axis. A photograph of a typical MnSb$_4$Te$_7$ crystal is shown in Figure 1a, with the dimensions of about 3 × 2 × 0.02 mm³. At ambient pressure, MnSb$_4$Te$_7$ crystal reveals a metallic behavior (Figure 1b). MnSb$_4$Te$_7$ exhibits an AFM transition at $T_N$ = 14.8 K[37], and correspondingly it leads to a kink in the resistivity $\rho(T)$ curve as shown in the inset of Figure 1b. AFM ordering temperature is further confirmed by the magnetic susceptibility presented in Figure 1c. The magnetic susceptibility decreases dramatically bellow $T_N$ for $H//c$, but only slightly decreases for $H//ab$, unveiling an anisotropic AFM exchange with the $c$ axis being the magnetic easy axis.

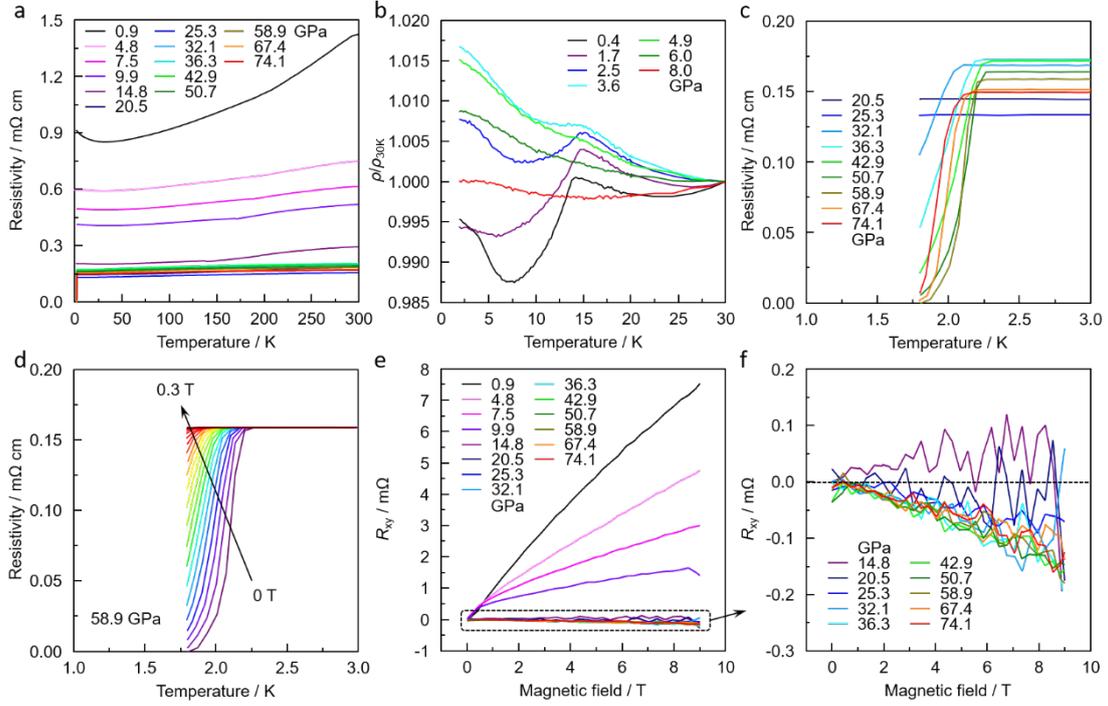

Figure 2. (a) Electrical resistivity of MnSb$_4$Te$_7$ as a function of temperature under high pressures in run III; (b) Temperature-dependent resistivity of MnSb$_4$Te$_7$ in the vicinity of the magnetic transition; (c) Temperature-dependent resistivity of MnSb$_4$Te$_7$ in the vicinity of the superconducting transition; (d) Temperature dependence of resistivity under different magnetic fields for MnSb$_4$Te$_7$ at 58.9 GPa; (e) and (f) Hall resistance of MnSb$_4$Te$_7$ as a function of magnetic field under various pressures at 10 K in run III.

Next, we measured the $\rho(T)$ of several MnSb$_4$Te$_7$ crystals under various pressures. Figure 2 shows the typical $\rho(T)$ curves of MnSb$_4$Te$_7$ up to 74.1 GPa. The sample exhibits metallic behavior in the whole pressure range. Increasing the pressure induces a continuous suppression of the overall magnitude of $\rho(T)$. It is well-known that AFM metallic ground state of MnBi$_2$Te$_4$ and MnBi$_4$Te$_7$ single crystal is gradually suppressed by pressure[29,35]. In order to obtained the trend of the AFM ordering temperature with pressure, a 400-micron culet DAC (BeCu) was used to produce a hydrostatic environment with NaCl as PTM. Figure 2b shown the detail of the normalized low-temperature resistivity of MnSb$_4$Te$_7$ as a function of temperature at various pressures to monitor the shift of the AFM transition kink. Unexpectedly, $T_N$ was insensitive to pressure until 8 GPa, which is notably different from MnBi$_2$Te$_4$ and MnBi$_4$Te$_7$[29,35]. Since the interlayer distance deceases under high pressure, it is speculated that the pressure-induced enhancement of AFM/FM competition and the partial delocalization of Mn-3$d$ electrons do not destroy long-range AFM order at low pressure region.

At a pressure of 32.1 GPa, a small resistivity drop appears at 1.9 K, indicating a superconducting phase transition (Figure 2c). Upon further increasing pressure, the critical temperature of superconductivity, $T_c$, gradually increases with pressure, and the maximum $T_c$ of 2.2 K is attained at $P$ = 50.7 GPa, as shown in Figure 2c. Beyond this pressure, $T_c$ decreases slowly, showing a dome-like behavior. Significantly, the $T_c$ of MnSb$_4$Te$_7$ is lower than that of Sb$_2$Te$_3$[13], which confirms the superconductivity is intrinsic and not derived from the latter. To gain further insight into the superconducting state, the temperature-dependent resistivity with an applied magnetic field was performed. As shown in Figure 2d, $T_c$ is suppressed progressively by magnetic fields, and a magnetic field of $\mu_0 H$ = 0.3 T deletes all signs of superconductivity above 1.8 K. Here, we determined $T_c$ as the 90% drop of the normal state resistivity. The $\mu_0 H_{c2}$ versus $T_c$ can be fitted well with Ginzburg-Landau formula: $\mu_0 H_{c2}(T) = \mu_0 H_{c2}(0)(1-t^2)/(1+t^2)$, where $t$ is the reduced temperature $T/T_c$. The obtained $\mu_0 H_{c2}(T)$ is 0.8 T and coherence length ξ is 20.3 nm (Figure S1).

MnBi$_4$Te$_7$ and MnSb$_4$Te$_7$ are sister compounds with the same structure (inset of Figure 1a) at ambient conditions. Both MnBi$_4$Te$_7$ and MnSb$_4$Te$_7$ show metallic behavior but the carrier type is opposite (electron-type for MnBi$_4$Te$_7$ and hole-type for MnSb$_4$Te$_7$). In order to trace the evolution of charge carriers upon compression, high-pressure Hall resistivity measurements was performed on MnSb$_4$Te$_7$ with pressure up to 74.1 GPa. Figures 2e and 2f display the Hall resistance curves $R_{xy}(H)$ under various pressures. At 0.9 GPa, the Hall resistance curve exhibits a linear feature with a positive slope, indicating a hole-type conduction in agreement with the previous reports[37]. With increasing the external pressure, the slope of the Hall resistance decreases dramatically and changes from positive to negative at 20.5 GPa, which implies a carrier-type inversion from hole- to electron-type. The carrier concentration is extracted from Hall measurements and the jump of carrier concentration coincides with the carrier-type inversion (Figure 4 and Figure S2). Above the critical pressure, the carrier concentration increases slowly.

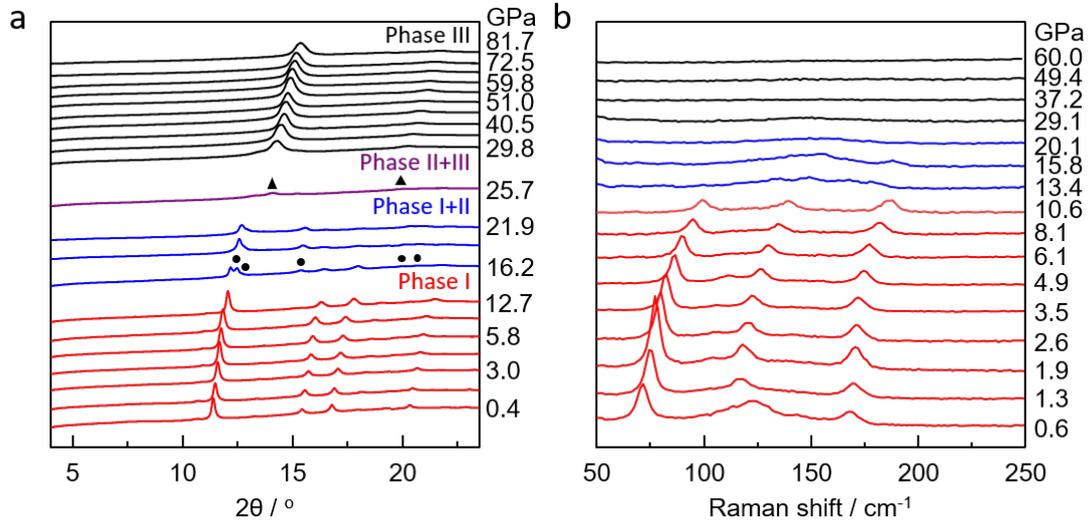

Figure 3. (a) XRD patterns of $MnSb_4Te_7$ measured at room temperature with increasing of external pressure up to 81.7 GPa. The x-ray diffraction wave-length $\lambda$ is 0.6199 Å. Red, blue, purple and black curves are used to distinguish the structure transformation; (b) Raman spectra at various pressures for $MnSb_4Te_7$ at room temperature.

In order to understand the underlying mechanism of the suppressed magnetism and induced superconductivity in $MnSb_4Te_7$ upon compression, we did an *in situ* study on structure evolutions by using synchrotron XRD and Raman spectroscopy, as presented in Figure 3. In the low-pressure range, all the diffraction peaks of $MnSb_4Te_7$ could be indexed to the rhombohedral $P\bar{3}m1$ structure, indicating robust ambient structure until $P \leq 12.7$ GPa (Figure S3). At 16.2 GPa, a high-pressure phase, phase II, was observed shown in Figure 3a. This phase is only stable in a narrow pressure range upon compression. Structure searches for this system by CALYPSO method[46-48] is failed since its complicated atom configuration. The Le Bail refinements for the high-pressure XRD data showed phase II attributed to monoclinic in space group $C2/m$. with $a =$ 14.2265(7) Å, $b = 3.8992(6)$ Å, $c = 16.7398(5)$ Å, and $\beta = 147.45(2)°$ at 21.9 GPa. On further increasing the pressure, $MnSb_4Te_7$ converted into a simple cubic phase with space group $Im\bar{3}m$ at 25.7 GPa. Interestingly, we obtained a pure phase III in the pressure range of 29.8 - 81.7 GPa. After a full pressure release, $MnSb_4Te_7$ recovers the ambient-pressure structure, indicating the reversible phase transition. The structural evolution of MnSb$_4$Te$_7$ under high pressure resembles the situation in the case of Sb$_2$Te$_3$[13]. The compression behavior is related to the distortion of MnTe$_6$ and SbTe$_6$ octahedra, which is similar to the close relative MnBi$_4$Te$_7$[35]. The pressure-induced structure evolution of MnSb$_4$Te$_7$ is also confirmed by *in situ* Raman spectroscopy measurements (Figure 3b). With increasing pressure, the profile of the spectra remains similar to that at ambient pressure until 13.4 GPa, then new vibration modes appear, thus showing the first structural transition. An abrupt disappearance of Raman peaks for pressure at 29.1 GPa indicates the structural transition to phase III. The evolution of the Raman spectra is consistent with the synchrotron XRD patterns. In a word, the results of synchrotron XRD together with Raman spectroscopy provide a clear evidence of pressure-induced structural transitions in MnSb$_4$Te$_7$.

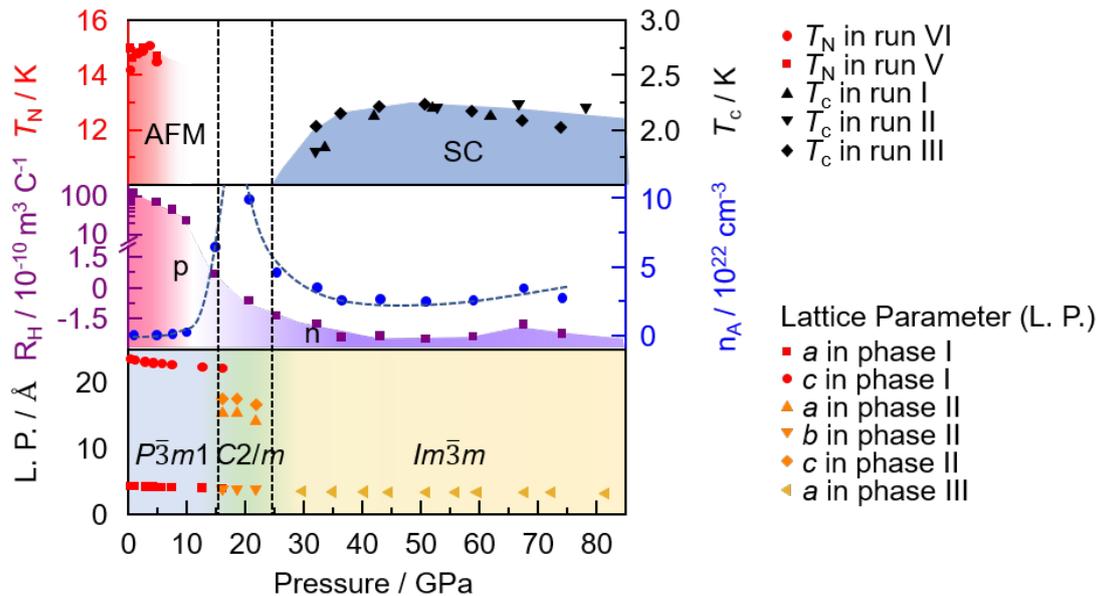

Figure 4. Phase diagram of MnSb$_4$Te$_7$. The upper panel shows the pressure dependence of $T_c$ and $T_N$. The middle panel shows the pressure dependence of Hall coefficient and carrier concentration at 10 K. The lower panel shows the lattice parameters as function of pressure.

The measurements on different samples of MnSb$_4$Te$_7$ provide the consistent and reproducible results, confirming that this superconductivity under pressure is intrinsic (Figure S4 and S5). The pressure dependences of the transition temperatures $T_N$/$T_c$, Hall resistance $R_{xy}(H)$, together with structural data obtained from the above measurements for MnSb$_4$Te$_7$ are mapped into the phase diagram shown in Figure 4. As a typical van der Waals material, the application of pressure induced two structure transitions in MnSb$_4$Te$_7$. For phase I ($P\bar{3}m1$), MnSb$_4$Te$_7$ show a typical AFM behavior with $T_N$ ~ 14.8 K. Interestingly, AFM ordering is robust at low pressure region, which is notably different from MnBi$_2$Te$_4$ and MnBi$_4$Te$_7$[35]. First structural transition appears at around 15 GPa accompanying with the elimination of AFM ordering. At the same time, the conduction changes from hole- to electron-type carrier, probable due to the reconstruction of Fermi surface. Phase II is only stable in a narrow pressure range and coexists with the phase I or the phase III upon compression. At around 25.7 GPa, MnSb$_4$Te$_7$ converted into a simple cubic phase, accompanying by the appearance of superconductivity. The $T_c$ increases with applied pressure and reaches a maximum value of 2.2 K at 50.7 GPa for MnSb$_4$Te$_7$, followed by a slow decrease, showing a dome shape phase diagram.

For MnSb$_4$Te$_7$ at ambient pressure, Mn $d$-orbitals have large contribution around the $E_F$[37]. Although atom coordination of Mn in phase III remains elusive, this large contribution of Mn $d$-orbital to density of states should not change significantly. The strong magnetism of Mn is commonly believed to be antagonistic to superconductivity. For a long time, manganese (Mn) is the only $3d$ element that does not show superconductivity among any Mn-based compounds, even though a great effort has been devoted recently to exploring the possible superconductivity via carrier doping or the application of high pressure. Recently, MnP was found to be the first Mn-based superconductor with transition temperature ~ 1 K under 8 GPa[49]. Another example is pressure-induced superconductivity in MnSe, where the interfacial effect between the metallic and insulating boundaries may play an important role[50]. Here in MnSb$_4$Te$_7$ system, superconductivity is observed accompanying with the structural transitions and suppression of AFM ordering, our results highlight that the possible antiferromagnetic spin fluctuations and pressure-induced cubic structure may be important for the appearance of superconductivity in MnSb$_4$Te$_7$. The mechanisms of pressure-induced structural and superconducting transitions in MnSb$_4$Te$_7$ deserve further studies.

In summary, we have performed a comprehensive high-pressure study on the electrical transport properties and crystal structures of the intrinsic MTIs MnSb$_4$Te$_7$. Application of pressure effectively tunes crystal structure of MnSb$_4$Te$_7$. Superconductivity with a dome shape behavior is observed for the first time after MnSb$_4$Te$_7$ converts into a simple cubic phase. Considering both intriguing magnetic topology and superconductivity in this material, our results call for further experimental and theoretical studies on MnSb$_4$Te$_7$ and related materials for a better understanding of the relation between magnetic and superconductivity, and its potential application in realizing topological superconductivity.

## ACKNOWLEDGMENT

We thank Prof. Hanyu Liu for valuable discussions. This work was supported by the National Natural Science Foundation of China (Grant No. 12004252, U1932217, 11974246), the National Key R&D Program of China (Grant No. 2018YFA0704300 and 2018YFE0202600), the Natural Science Foundation of Shanghai (Grant No. 19ZR1477300), the Science and Technology Commission of Shanghai Municipality (Grant No. 19JC1413900), Shanghai Science and Technology Plan (Grant No. 21DZ2260400) and the Beijing Natural Science Foundation (Grant No. Z200005). The authors thank the support from Analytical Instrumentation Center (# SPST-AIC10112914), SPST, ShanghaiTech University. The authors thank the staffs from BL15U1 at Shanghai Synchrotron Radiation Facility for assistance during data collection.